\documentclass[nohyperref]{article}

\usepackage{microtype}
\usepackage{graphicx}
\usepackage{subfigure}
\usepackage{booktabs} 

\usepackage{hyperref}

\usepackage[accepted]{icml2022}

\usepackage{amsmath}
\usepackage{amssymb}
\usepackage{mathtools}
\usepackage{amsthm}

\usepackage[capitalize,noabbrev]{cleveref}

\newcommand{\snn}{{\sc SuperNNova }}
\newcommand{\ssnn}{{\sc SNN }}
\newcommand{\ssnns}{{\sc SNN}}

\usepackage[textsize=tiny]{todonotes}

\icmltitlerunning{Bayesian Neural Networks for classification tasks in the Rubin big data era}

\begin{document}

\twocolumn[
\icmltitle{Bayesian Neural Networks for classification tasks in the Rubin big data era}




\begin{icmlauthorlist}
\icmlauthor{Anais M{\"o}ller}{SUT}
\icmlauthor{Thibault de Boissi{\`e}re}{Canva}
\end{icmlauthorlist}

\icmlaffiliation{SUT}{Centre for Astrophysics and Supercomputing, Swinburne University of Technology, Mail Number H29, PO Box 218, 31122 Hawthorn, VIC, Australia.}
\icmlaffiliation{Canva}{Canva inc., Sydney, Australia}

\icmlcorrespondingauthor{Anais M{\"o}ller}{amoller@swin.edu.au}

\icmlkeywords{Machine Learning, ICML}

\vskip 0.3in
]



\printAffiliationsAndNotice{}  

\begin{abstract}
Upcoming surveys such as the Vera C. Rubin Observatory Legacy Survey of Space and Time (LSST) will detect up to 10 million time-varying sources in the sky every night for ten years. This information will be transmitted in a continuous stream to brokers that will select the most promising events for a variety of science cases using machine learning algorithms. We study the benefits and challenges of Bayesian Neural Networks (BNNs) for this type of classification tasks. BNNs are found to be accurate classifiers which also provide additional information: they quantify the classification uncertainty which can be harnessed to analyse this upcoming data avalanche more efficiently.
\end{abstract}

\section{Introduction}

We are entering into a new era of big data time-domain astronomy. Upcoming surveys such as Vera C. Rubin Observatory Legacy Survey of Space Time (LSST) will detect up to ten million time-domain events every night for over a decade. LSST will emit an alert stream with time-domain even data within minutes of observation. The Rubin Community brokers will then receive that stream in real-time. Brokers will enrich and filter these alerts to select the most promising candidates for a variety of science cases. The selected LSST brokers are {\sc ALeRCE} \citep{Forster:2020}, {\sc Ampel} \citep{ampel}, {\sc Babamul}, {\sc Antares} \citep{antares}, {\sc Fink} \citep{Moller:2021}, {\sc Lasair} \citep{Smith_2019} and  {\sc Pitt-Google}. Many of these brokers are currently processing the Zwicky Transient Facility (ZTF) alert stream which is an order of magnitude less than what is expected from Rubin. 

Classification algorithms are a core part of brokers. They provide scores that can be used to select candidates for specific astrophysical phenomena such as supernovae, kilonovae, microlensing, AGNs, and many others. In recent years, a wide variety of classification algorithms have been developed for time-domain astronomy which have shown excellent performance in classification tasks \cite{Leoni:2021,Muthukrishna2019,Villar:2019,Villar:2020,Godines2019,Ishida:2019} 

Bayesian Neural Networks are a promising classification method that provide classification scores as well as uncertainties that can reflect the model’s confidence in the prediction. BNNs go beyond point estimates and yield a distribution of classification probabilities. The final prediction is typically computed as the mean probability of this distribution. The standard deviation of this distribution is typically used as an estimation of the prediction uncertainty. The astronomical community has recently started to use BNNs for classification tasks \cite{Walmsley:2020,Moller:2020,Moller:2022}.

In this work, we use the \snn (\ssnns) classification algorithm \cite{Moller:2020} to evaluate the performance and interpretability of BNNs for the new Rubin era. Since the Rubin observing strategy is yet to be defined, we use the Dark Energy Survey (DES) SN fields as a proxy for Rubin's Deep Drilling Fields. Our benchmark task is the classification of type Ia vs non Ia supernovae light-curves. Supernovae are bright stellar explosions that fade away within weeks. 

\begin{figure}[ht!]
\vskip 0.2in
\begin{center}
\centerline{\includegraphics[width=\columnwidth]{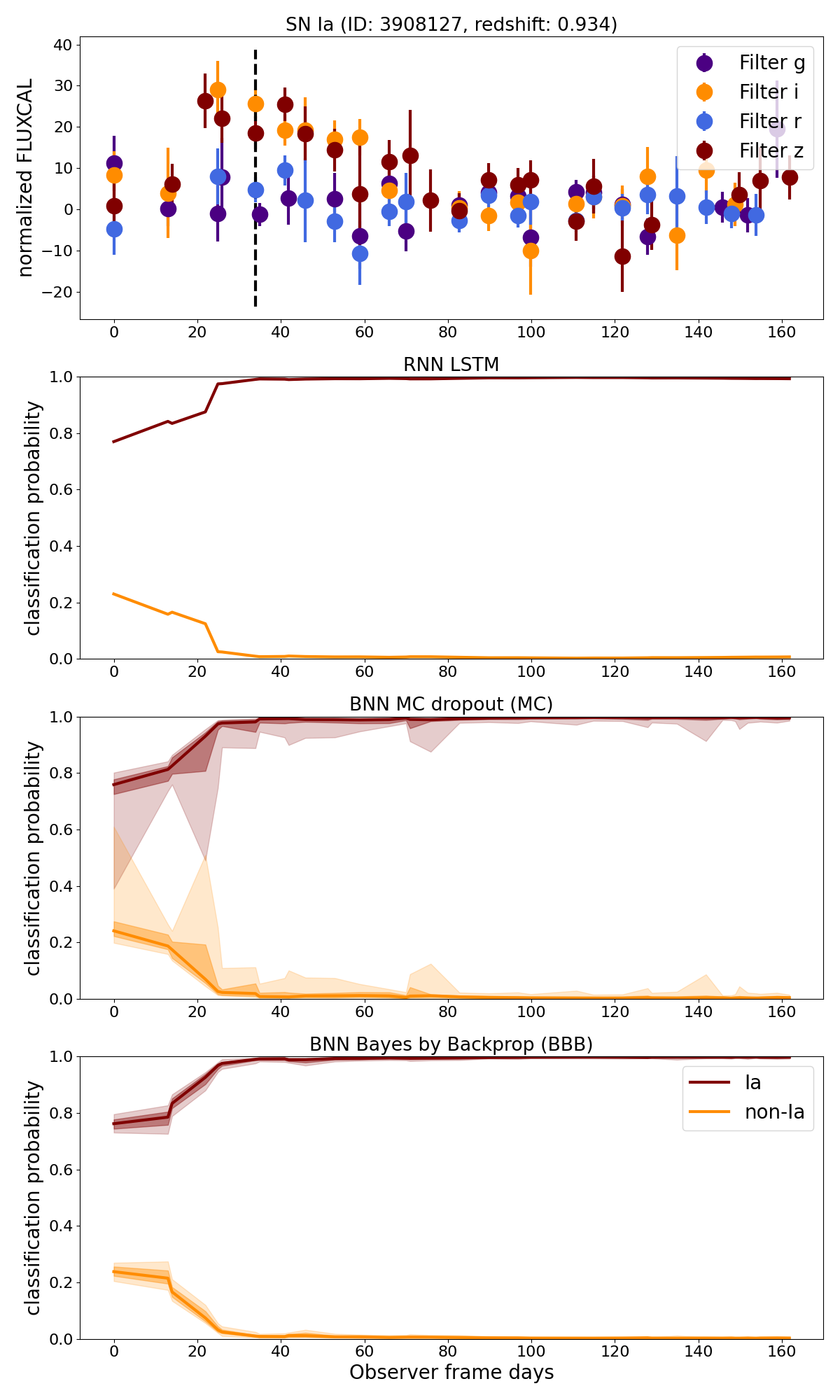}}
\caption{Simulated DES light-curve and SNN classifications using three different neural networks: RNN, BNN MC, and BNN BBB. Top row shows the SN light-curve (normalised flux in band-passes $g,r,i,z$; time in Observer Frame days) and its time of maximum brightness in a dashed line. Bottom rows shows the classification scores for each method (SN Ia: maroon, non SN Ia: orange). The BNN methods provide classification uncertainties (shadowed regions show 68 per cent and 95 per cent contours).}
\label{fig:lc}
\end{center}
\end{figure}

We use a simulated dataset for training and evaluation. These simulations contain realistic DES light-curves from Type Ia models, peculiar Ia and core-collapse supernovae. An example of the simulated light-curves can be seen in Figure~\ref{fig:lc}. More details on these simulations can be found in \cite{Moller:2022}. Additionally, we use real light-curves of type Ia supernovae candidates observed by the Zwicky Transient Facility (ZTF) for evaluation in Section~\ref{sec:ZTF}. These light-curves where obtained using the Fink broker API\footnote{\url{https://fink-portal.org/api}}.

\subsection{Bayesian Neural Networks (BNNs)}
Scientific application of machine learning methods often require estimating the model's predictive uncertainty. A popular way to do so is to cast NN training under a bayesian light where the goal is to learn a probability distribution over possible neural networks. This problem is typically untractable analytically. In practice, approximate methods, such as variational inference, are used. A review of BNNs and their use in astronomy can be found in \cite{Charnock:2020}.

We use two BNN implementations to approximate the posterior distribution of weights: MC dropout \citep[MC;][]{Gal:2015} and Bayes by Backprop \citep[BBB;][]{Fortunato:2017}. MC provides a Bayesian interpretation of recurrent dropout when the dropout mask is the same at all time steps. BBB learns an approximate posterior distribution of weights using variational inference. Both methods have been previously implemented and tested on simulations in \snn \citep{Moller:2020}.

For both methods, to obtain the classification probability distribution, we sample the predictions from our BNN 50 times. In the following we compute the classification probability, $\widehat p_i$ for a given light-curve, $\mathbf{x}_i$ as:
\begin{equation}\label{eq:BNN_probability}
\widehat p_i = mean \bigg\{ p_{j}({\mathbf{x}_i})\bigg\}_{i=1:n_s}
\end{equation}
where $j \in [1,n_s]$ is the index of inference samples, $p_{j}({\mathbf{x}_i})$ is the $j$-th sample of the classification probability distribution for the light-curve $\mathbf{x}_i$.

To evaluate classification uncertainties we compute the model uncertainty for a given light-curve $\mathbf{x}_i$ as:
\begin{equation}\label{eq:uncertainty}
\widehat \sigma_i = std \bigg\{ p_{j}({\mathbf{x}_i})\bigg\}_{i=1:n_s}
\end{equation}
where $std$ is the sample standard deviation. 

\section{Calibration for BNNs}
Classification probabilities that reflect the real likelihood of events being correctly assigned to a target are said to be calibrated. Calibration is extremely important in the Rubin context to carry out precision cosmology analyses.

\begin{figure}[ht]
\vskip 0.2in
\begin{center}
\centerline{\includegraphics[width=\columnwidth]{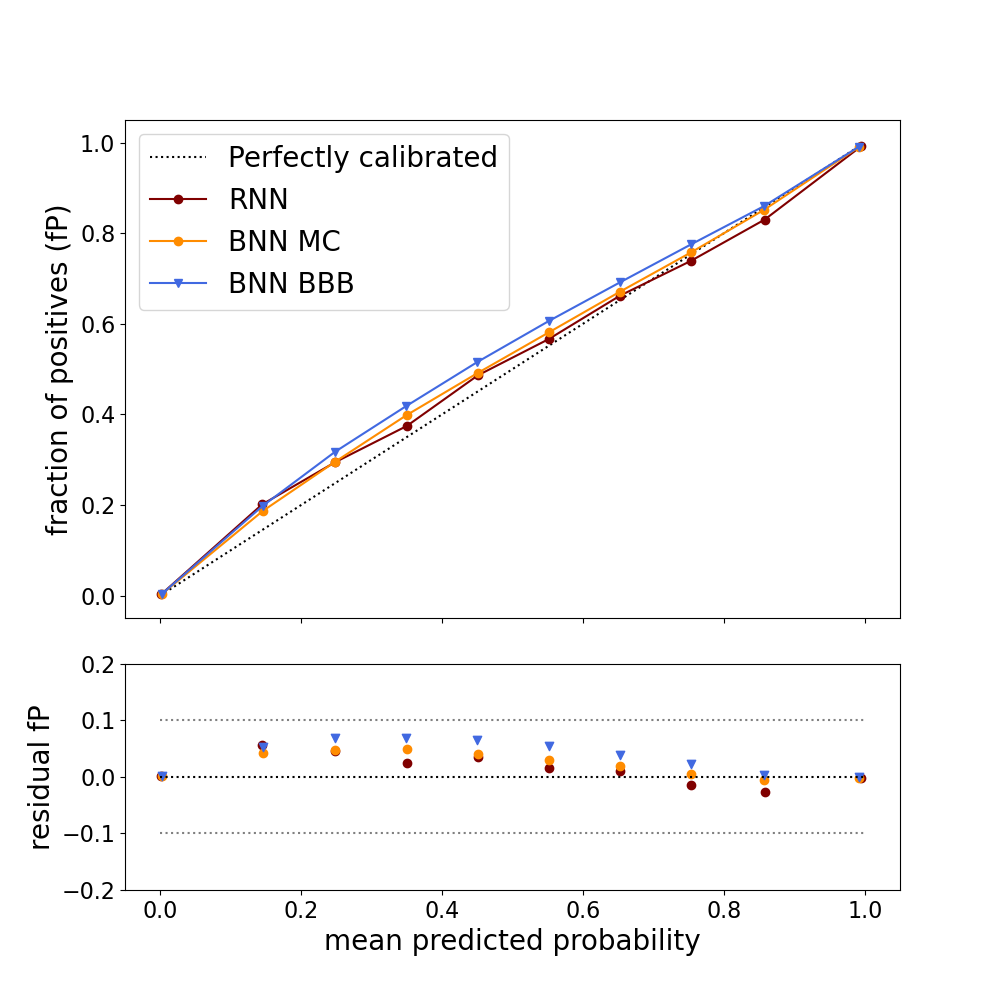}}
\caption{Calibration: reliability diagrams for an RNN and two BNNs: MC dropout (MC) and Bayes by Backprop (BBB).}
\label{fig:calibration}
\end{center}
\vskip -0.2in
\end{figure}

We use reliability diagrams \cite{DeGroot:1983} to evaluate our model's calibration in Figure~\ref{fig:calibration}. We evaluate this calibration with complete light-curves spanning $\approx 100-150$ days. We find all models to be close to perfectly calibrated with some excess on the fraction of positives at low probability in particular for BNN BBB. 

\section{Early classification accuracy}
To identify promising time-domain events swiftly, it is necessary to use algorithms that allow accurate classification with only a couple of observations. This is increasingly necessary as surveys such as Rubin will detect millions of transients per night which need to be disentangled together with the scarce follow-up resources which need to be optimized.

\begin{figure}[ht]
\vskip 0.2in
\begin{center}
\centerline{\includegraphics[width=\columnwidth]{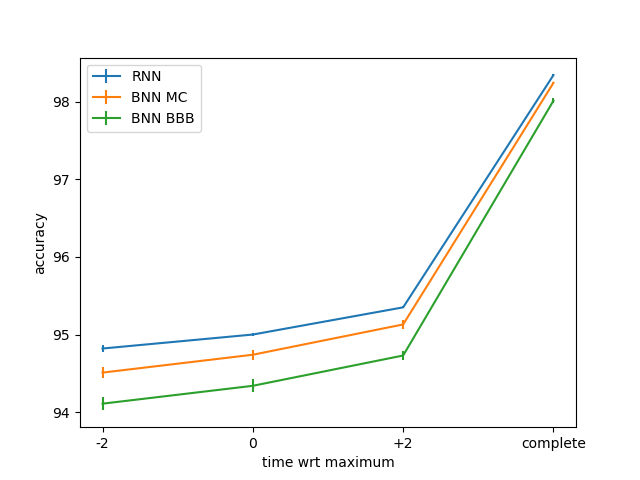}}
\caption{Accuracy of type Ia vs. non Ia supernova classification with respect to time of maximum brightness in the rest frame of a supernova. We evaluate an RNN and two BNNs: MC dropout (MC) and Bayes by Backprop (BBB) for the same task. Experiments are repeated 5 times with different random seeds. The reported accuracy is the mean accuracy. The reported error bar is the standard deviation.}
\label{fig:acc_time}
\end{center}
\vskip -0.2in
\end{figure}

We evaluate the performance of a traditional RNN and the two BNNs in \ssnn for early classification (around maximum brightness) in Figure~\ref{fig:acc_time}. We consider a light-curve classified as type Ia if its classification probability is $\widehat p_{Ia}>0.5$. Early classification is found to be highly-accurate. As more observations become available we also find that the accuracy increases. We note that BNNs are shown to have slightly lower accuracies than a LSTM RNN. This may be improved with careful tuning of the BNN prior parameters. 

In the following, we continue exploring BNNs in this classification task with complete light-curves.

\section{Interpretability: Out of distribution (OOD) or anomalies}
In this Section we present two tests designed to evaluate the interpretability of BNN predictions for OOD or anomalies.

\subsection{Entropy}
Entropy has been used as a proxy for the model's confidence in its predictions and thus an interesting metric to evaluate BNNs on \cite{Fortunato:2017}. Confident predictions should yield low entropy.
For a dataset $\mathcal{D}:[\mathbf{x}_1,...\mathbf{x}_N]$ with $N$ light-curves and a classification model $m$, the entropy of $\mathcal{D}$ under $m$ is defined as:
\begin{equation}
H_m[\mathcal{D}] = \sum_{i=1}^N p(\mathbf{y}_i|\mathbf{x}_i) log \left( \frac{1}{p(\mathbf{y}_i|\mathbf{x}_i)} \right).
\end{equation} \label{eq:entropy}
where $p(\mathbf{y}_i|\mathbf{x}_i)$ is the classification probability given the light-curve $i$.

For two given sets of predictions, we can define the {\it entropy gap} $\Delta H$ by:
\begin{equation}\label{eq:delta_entropy}
\Delta H = \quad \overline{H_{m1}}[\mathcal{D}] - \overline{H_{m2}}[\mathcal{D}] \quad 
\end{equation}
where we evaluate the entropy gap over the same dataset for two given models ($m_1,m_2$). It is expected that predictions on OOD datasets would have a large entropy if the algorithm is robust. Thus the $\Delta H$ between these OOD and a SN-like dataset should be positive.

We generate three different types of OOD events: time reversed light-curves, randomly shuffled light-curves and random fluxes. We evaluate the entropy of these predictions in Table~\ref{tab:entropy}. For reverse and shuffled light-curves we find, as expected, a high entropy gap when compared to supernova light-curve classification. This behaviour is not seen in random light-curves which may be attributed to their possible resemblance to noisy low signal-to-noise supernovae.

\begin{table}[t]
\caption{Delta entropy between out-of-distribution events and supernova classification.}
\label{tab:entropy}
\vskip 0.15in
\begin{center}
\begin{small}
\begin{sc}
\begin{tabular}{lcccr}
\toprule
model & Random & Reverse & Shuffle\\
\midrule
RNN & -0.02 & 0.01 & 0.03 \\
MC & -0.02 & 0.05 & 0.1 \\
BBB & -0.02 & 0.08 & 0.06\\
\bottomrule
\end{tabular}
\end{sc}
\end{small}
\end{center}
\vskip -0.1in
\end{table}

\subsection{DES-like vs. ZTF light-curves}\label{sec:ZTF}
We now explore the predictions obtained for a DES-like dataset, which is similar to the training set, and SNe Ia candidates from ZTF selected by the {\sc Fink} broker. These candidates have been selected as probable SNe Ia by classifiers such as \snn trained on ZTF-like data.

MC models (resp. BBB) trained on DES-like simulations of type Ia supernovae obtain on average a median classification probability of 0.9 (resp. 0.9) and standard deviation of the classification probability distribution of 0.002 (resp. 0.001 for BBB). Contrast this with an average classification probability of 0.6 (resp. 0.6) and standard deviation of 0.19 (resp. 0.27) on the early SNe Ia candidates from the Fink broker. Clearly, the models are less confident in their prediction for this new dataset showing an expected interpretable behaviour.

\section{Scalability}

As Rubin's data volume will be unprecedented, we require fast classification algorithms. 
\ssnn RNN has been bench-marked in the {\sc Fink} broker, classifying 2500 light-curves per second/core. 

Here, we use a simpler benchmark to assess the classification accuracy of BNNs with respect of the number of samples of their probability distribution. We evaluate the classification of 1000 light-curve using one node and the embedded \ssnn database.

In Table~\ref{tab:acc_numinf} we show the classification accuracy evolution as a function of number of samples of the probability distribution. We find the decrease in accuracy is small ($<0.2$ for BBB while $<0.01$ for MC) compared to the reduction of classification time in the algorithms (up to one order of magnitude). Thus, for brokers it could be envisaged to reduce the number of samples to provide fast classification with these algorithms. We note that an evaluation on the impact of this sampling in the estimated uncertainties is left for future work.

\begin{table}[t]
\caption{Model classification accuracy, duration of classification for 1000 light-curves and number of samples used for the results.}
\label{tab:acc_numinf}
\vskip 0.15in
\begin{center}
\begin{small}
\begin{sc}
\begin{tabular}{lcccr}
\toprule
model & accuracy & time(s) & number sampling \\
\midrule
RNN & 98.29 & 10 & NA \\
\hline
BBB & 98.09 & 315 & 50 \\
BBB & 97.89 & 129 & 20 \\
BBB & 97.89 & 35 & 5 \\
\hline
MC & 98.39 & 342 & 50 \\
MC & 98.39 & 142 & 20 \\
MC & 98.39 & 39 & 5 \\
\bottomrule
\end{tabular}
\end{sc}
\end{small}
\end{center}
\vskip -0.1in
\end{table}

\section{Summary}
The use of Bayesian Neural Networks (BNNs) for classification has began within the astronomical community. In this work, we explore the use of BNNs in classification tasks. We performed calibration and scalability benchmarks and explored the interpretability of BNNs outputs. 

BNNs are found to be highly accurate in classification tasks with calibrated probabilities. We have also found quantitative evidence that the prediction confidence of our BNNs decreases for out-of-distribution events.

With the advent of large surveys discovering thousands of transients every night, it will be imperative to prioritize follow-up using partial light-curve classification. {\sc SuperNNova} BNNs achieve high accuracy on this challenging task. Future work will expand on the study of robustness of BNNs, in particular 
supplementing probability thresholds with prediction uncertainties to improve classification.

\section*{Software and Data}

Software used in this work is open source and available in GitHub. Trained models are available on request.

\section*{Acknowledgements}
We thank the anonymous reviewers for feedback to improve this work. 

This work was developed within the Fink community and made use of the Fink community broker resources. Fink is supported by LSST-France and CNRS/IN2P3.


\nocite{langley00}

\bibliography{example_paper}
\bibliographystyle{icml2022}



\end{document}